\documentclass[12pt,a4paper]{article}

\usepackage{amsmath}
\usepackage{graphicx}

\makeatletter
\def\Ginclude@eps#1{%
 \message{<#1>}%
  \bgroup
  \def\@tempa{!}%
  \dimen@\Gin@req@width
  \dimen@ii.1bp%
  \divide\dimen@\dimen@ii
  \@tempdima\Gin@req@height
  \divide\@tempdima\dimen@ii
    \includegraphics{#1}%
  \egroup}
\makeatother

\begin{document}

\title{A derivation of the Doppler factor in the \\ Li\'{e}nard-Wiechert potentials}

\author{C\u alin Galeriu}

\maketitle

\section*{Abstract}

We present an elegant derivation of the
Doppler factor in the Li\'{e}nard-Wiechert potentials of an extended particle,
based on a theorem authored by J.~L.~Synge and
dealing with cross-sections through worldtubes. 

\section*{}

The Doppler factor in the Li\'{e}nard-Wiechert (LW) potentials of an electrically
charged extended particle arises from the evaluation of the electric charge density 
at the retarded time.
Starting with Li\'{e}nard and Wiechert, many
authors have derived this Doppler factor one way or another, see for example \cite{Heras},
\cite{galeriuArXiv21},
and the references therein.
For an extended particle of very small size and uniform electric charge density $\rho$, 
the actual electric charge at time $t$ is
\begin{equation}
Q_{actual} = \int \rho(\overrightarrow{r}, t) \ dx \ dy \ dz = \rho\ V_{actual}, 
\label{eq:1}
\end{equation}
and the apparent electric charge, relative to a given field point, is
\begin{equation}
Q_{apparent} = \int \rho(\overrightarrow{r_{ret}}, t_{ret}) \ dx \ dy \ dz = \rho\ V_{apparent},
\label{eq:2}
\end{equation} 
where $V_{actual}$ is the actual volume occupied by the particle at time $t$
and $V_{apparent}$ is the apparent volume, the volume of the integration domain
where the retarded electric charge density is not zero.
The Doppler factor in the LW potentials is the ratio
\begin{equation}
\frac{Q_{apparent}}{Q_{actual}} = \frac{V_{apparent}}{V_{actual}} = \frac{1}{1 - v_r / c},
\label{eq:3}
\end{equation}
where $v_r$ is the radial component of the retarded velocity,
and $c$ is the speed of light in vacuum.

\begin{figure}[h!]
\centering
\includegraphics[height=13.9347cm]{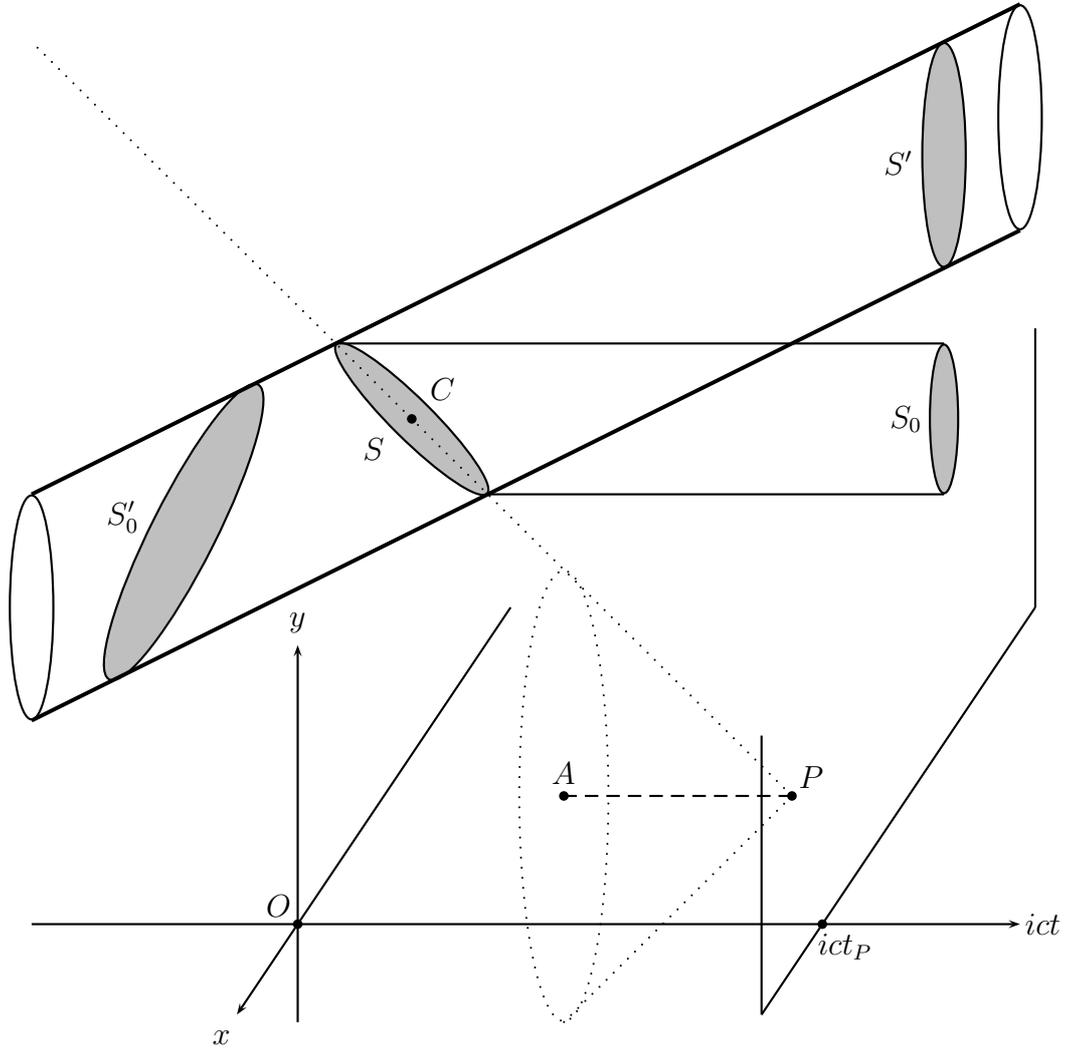}
\caption{The actual volume is the volume of the intersection $S'$
of the particle's worldtube with the hyperplane of constant time $t = t_P$.
The apparent volume is the volume of the projection $S_0$ on the same hyperplane
of the intersection $S$ of the particle's worldtube with the retarded lightcone
drawn through the field point $P$.}
\label{fig:1}
\end{figure}

Consider an electrically charged extended particle moving with 
velocity $\overrightarrow{v} = ( v_x , v_y , v_z )$. 
In Minkowski space the particle is represented by a worldtube, as seen in Fig.~\ref{fig:1}. 
Consider also the field point $P$ at $(x_P, y_P, z_P, i c t_P)$.
The use of an imaginary time coordinate, mathematically equivalent to 
the more modern use of a Minkowski metric tensor, brings us closer to the familiar Euclidean vectors.
\lq\lq Although this imaginary coordinate makes some formulae harder to interpret physically,
this disadvantage is outweighed by the fact we do not have to distinguish between covariant
and contravariant components, and all suffixes may be written as subscripts.\rq\rq \cite{synge1970}
Through the field point $P$ we draw the hyperplane of constant time $t = t_P$.
The intersection of this hyperplane with the 
worldtube of the particle gives the region labeled $S'$, whose volume $S'$ is the actual volume. 
Through the field point $P$ we also draw the retarded lightcone. 
The intersection of this lightcone
with the worldtube of the particle, the region labeled $S$,
is subsequently projected on the hyperplane of constant time $t = t_P$, 
and gives the region labeled $S_0$, whose volume $S_0$ is the apparent volume. 
Relative to the center $C$ of the retarded electric charge density,
the field point $P$ is described by a position vector
$\overrightarrow{r} = ( r_x , r_y , r_z ) = ( x_P - x_C , y_P - y_C , z_P - z_C )$.
The radial component of the retarded velocity is $v_r = \overrightarrow{v} \cdot \overrightarrow{r} / r$.

Looking at Fig.~\ref{fig:1}, 
we realize that both the actual volume $S'$ and the apparent volume $S_0$ are
the volumes of
cross-sections through worldtubes. Indeed, in addition to the worldtube of the
electrically charged particle, we also have a worldtube parallel to the time axis,
graphically generated by the projection of the intersection region $S$
(the region with non-zero retarded electric charge density)
on the Euclidean space of the field point $P$ (the 3D space at time $t = t_P$). 
It therefore seems natural to investigate whether the relationship between the
apparent and actual volumes of a particle, given in Eq. (\ref{eq:3}), could be derived 
based on Synge's theorem about cross-sections through worldtubes. \cite{synge}

Synge's worldtube theorem states that 
\begin{equation}
S_0 = S \left|\overrightarrow{\lambda} \cdot \overrightarrow{n}\right|,
\label{eq:4}
\end{equation}
where $S_0$ is the volume of the normal cross-section of a worldtube,
$S$ is the volume of an oblique cross-section with unit normal $\overrightarrow{n}$,
and $\overrightarrow{\lambda}$ is a unit vector tangent to the worldtube.
As Synge noticed, Eq. (\ref{eq:4}) is the Minkowski space analogue of the 
formula $S_0 = S \cos(\theta)$ for the projection of an area in 3D Euclidean space.

In order to be able to apply this theorem, first we notice that, for an extended particle
of finite size, the intersection region $S$ is not a cross-section. This fact, however, 
is of no concern to us, since the calculation of the LW potentials is done for an extended particle
of infinitesimal size. Locally, around the center $C$ of the retarded electric charge density,
the curved surface of the lightcone can be approximated by a plane (hyperplane) surface.
In a similar manner Aguirregabiria {\it et al.} \cite{Aguirregabiria} have approximated the 
information collecting sphere 
(the intersection of the 3D space with the retarded lightcone, at a given time)
with an "information collecting plane" when calculating 
the retarded shape of an infinitesimally small moving sphere.

Let $S'_0$ be the volume of a normal cross-section through the worldtube of the particle.
A straightforward application of Synge's worldtube theorem seems to produce three equations
\begin{eqnarray}
S_0 = S \left|\overrightarrow{\lambda} \cdot \overrightarrow{n}\right|, \label{eq:5} \\
S'_0 = S \left|\overrightarrow{\lambda'} \cdot \overrightarrow{n}\right|, \label{eq:6} \\
S'_0 = S' \left|\overrightarrow{\lambda'} \cdot \overrightarrow{n'}\right|, \label{eq:7}
\end{eqnarray}
from which one could extract the Doppler factor as
\begin{equation}
\frac{S_0}{S'} = \frac{\left|\overrightarrow{\lambda'} \cdot \overrightarrow{n'}\right|
\ \left|\overrightarrow{\lambda} \cdot \overrightarrow{n}\right|}
{\left|\overrightarrow{\lambda'} \cdot \overrightarrow{n}\right|}.
\label{eq:8}
\end{equation}
Unfortunately, we are now faced with a difficulty. 
While the unit vector tangent to the worldtube
of the particle
\begin{equation}
\overrightarrow{\lambda'} = \frac{\gamma}{c} ( v_x , v_y , v_z , i c ) 
= \frac{\gamma}{c} ( \overrightarrow{v} , i c ),
\label{eq:9}
\end{equation}
the unit vector tangent to the worldtube generated by the projection of $S$
\begin{equation}
\overrightarrow{\lambda} = ( 0 , 0 , 0 , i ),
\label{eq:10}
\end{equation}
and the unit normal to the cross-section $S'$
\begin{equation}
\overrightarrow{n'}= ( 0 , 0 , 0 , i ),
\label{eq:11}
\end{equation}
are well defined quantities, the unit normal to the cross-section $S$, $\overrightarrow{n}$, is not.

A similar problem was encountered by Synge when he tried to apply his theorem to a null
worldtube, a worldtube in the direction of a light signal. In this special case
no unit vector tangent to the worldtube and no normal cross-section do exist.
His solution was to approach the special case as a limit, first by writing the ratio of the
equations for two different oblique cross-sections of a non-null worldtube
\begin{equation}
\frac{S_0}{S_0} = \frac{S_1 \left|\overrightarrow{\lambda} \cdot \overrightarrow{n_1}\right|}
{S_2 \left|\overrightarrow{\lambda} \cdot \overrightarrow{n_2}\right|},
\label{eq:12}
\end{equation}
next by replacing the tangent unit vector $\overrightarrow{\lambda}$ with
a tangent vector of unspecified length $\overrightarrow{t}$, 
and finally by changing the direction of the worldtube
into a null direction. In this way the unusable formula (\ref{eq:4}) turns into
\begin{equation}
S_1 \left|\overrightarrow{t} \cdot \overrightarrow{n_1}\right|
= S_2 \left|\overrightarrow{t} \cdot \overrightarrow{n_2}\right|,
\label{eq:13}
\end{equation}
where $S_1$ and $S_2$ are the volumes of two oblique cross-sections,
$\overrightarrow{n_1}$ and $\overrightarrow{n_2}$ are the unit 
normals to these cross-sections, and $\overrightarrow{t}$ is any vector (of null length) tangent 
to the null worldtube.
As Synge noticed, Eq. (\ref{eq:13}) is the Minkowski space analogue of the 
formula $S_1 \cos(\theta_1) = S_2 \cos(\theta_2)$ from 3D Euclidean space.

We follow the same approach here. We don't draw the retarded lightcone 
through the field point $P$,
but instead we draw a retarded hypercone of smaller apex angle, 
a hypersurface described by the equation
\begin{equation}
( x - x_P )^2 + ( y - y_P )^2 + ( z - z_P )^2 - w^2 ( t - t_P )^2 = 0,
\label{eq:14}
\end{equation}
where $t < t_P$ and $w < c$. For this hypercone we can write Eq. (\ref{eq:8}) without any issues, and
then we replace 
the unit normal $\overrightarrow{n}$ with a normal vector $\overrightarrow{m}$ of any length. 
In this way we obtain
\begin{equation}
\frac{S_0}{S'} = \frac{\left|\overrightarrow{\lambda'} \cdot \overrightarrow{n'}\right|
\ \left|\overrightarrow{\lambda} \cdot \overrightarrow{m}\right|}
{\left|\overrightarrow{\lambda'} \cdot \overrightarrow{m}\right|},
\label{eq:15}
\end{equation}
where $\overrightarrow{m}$ is any vector normal to the cross-section $S$. Equation (\ref{eq:15}) can
be used with vectors $\overrightarrow{m}$ of any length, including vectors of null length. 
The hypercone (\ref{eq:14}) turns into a lightcone
in the limit $w \to c$, and we have to discover what happens to the normal $\overrightarrow{m}$
in this special situation.

Consider a right circular cone in 3D Euclidean space, and a point somewhere on the curved surface of this cone. Through this point we draw the normal to the surface. 
From pure symmetry considerations we conclude that the normal to the
surface, the line connecting the given point to the vertex of the cone, and the 
symmetry axis of the cone are all coplanar. The cone has circular symmetry,
and the plane of interest is a mirror plane.

\begin{figure}[h!]
\centering
\includegraphics[height=8.1844cm]{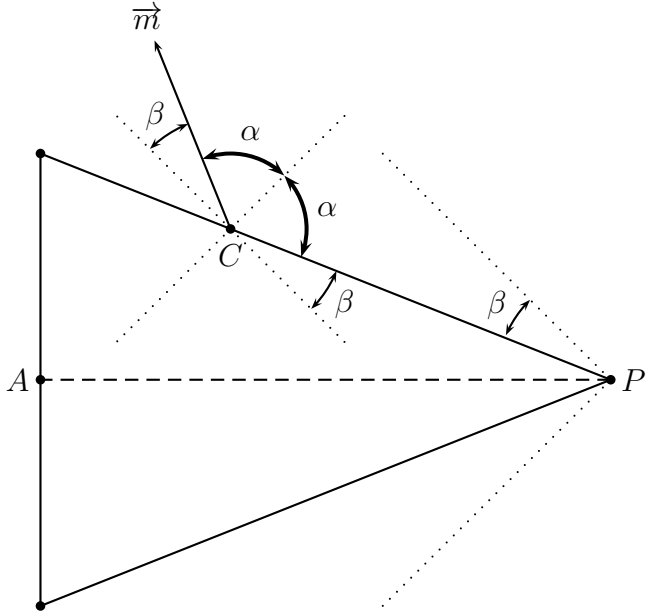}
\caption{The Euclidean representation of the Minkowski plane 
that contains the normal to the hypercone's surface in point $C$ and 
 the hypercone's symmetry axis $AP$.}
\label{fig:2}
\end{figure}

The same symmetry argument applies in the case of a hypercone in 4D Minkowski space.
As seen in Fig.~\ref{fig:2}, the normal to the hypercone's surface in point $C$, 
the line connecting point $C$ on the surface to vertex $P$, and 
the hypercone's symmetry axis $AP$ are all coplanar.

We also know that the normal $\overrightarrow{m}$ is perpendicular to line $CP$.
In the Euclidean representation of the Minkowski plane, this orthogonality condition
manifests itself as the congruence of the angles made by $\overrightarrow{m}$
and $CP$ with the bisector of the first quadrant. We name these angles $\alpha$.
The angle made by $\overrightarrow{m}$ with the bisector of the second quadrant
and the angle made by $CP$ with the bisector of the fourth quadrant 
are also congruent. We name these angles $\beta$.
 
In the limit $w \to c$, when the hypercone turns into a lightcone,
the limit $\beta \to 0$ is also approached at the same time,
and both the normal $\overrightarrow{m}$ and line $PC$ 
(connecting the field point $P$ to the center $C$ of the retarded electric charge density)
take the direction of the bisector of the second quadrant, thus becoming parallel.

Since $C$ and $P$ are connected by a light signal, we know that
\begin{equation}
\overrightarrow{CP} = ( x_P - x_C , y_P - y_C , z_P - z_C , i c t_P - i c t_C ) = ( \overrightarrow{r} , i r ),
\label{eq:16}
\end{equation}
and since $\overrightarrow{m}$ can be any vector normal to the cross-section $S$, we can choose
\begin{equation}
\overrightarrow{m} = \overrightarrow{PC} = ( - \overrightarrow{r}, - i r).
\label{eq:17}
\end{equation}

In the final step of our derivation, direct substitution of Eqs. (\ref{eq:9}), (\ref{eq:10}), (\ref{eq:11}), 
and (\ref{eq:17}) into Eq. (\ref{eq:15}) produces the expected Doppler factor (\ref{eq:3}).

In conclusion, the Doppler factor in the LW potentials
of an electrically charged extended particle
is fully determined by the relationship between different 
cross-sections through worldtubes.
The elegant derivation presented here,
based on Synge's worldtube theorem,
also reminds us that this Doppler factor
\lq\lq is a purely {\it geometrical} effect.\rq\rq \cite{Griffiths4ed} 
One should also keep in mind that this derivation
works only for extended electric charges, and cannot be used for electric point charges.
This is because in Minkowski space an extended charge is represented by
a worldtube, which has a cross-section, while a point charge is represented
by a worldline, which has no cross-section, the intersection between
a worldline and a hyperplane being just a point.


\begin{thebibliography}{9}

\bibitem{Heras} Ricardo~Heras, 
\lq\lq Alternative routes to the retarded potentials\rq\rq,
{\it Eur. J. Phys.} {\bf 38} (5), 055203 1-13 (2017).

\bibitem{galeriuArXiv21} C\u alin~Galeriu, 
\lq\lq The geometrical origin of the Doppler factor in the Li\'{e}nard-Wiechert potentials\rq\rq, 
arXiv:2101.12337 [physics.class-ph] (2021).

\bibitem{synge1970} J.~L.~Synge, 
\lq\lq Point-particles and energy tensors in special relativity\rq\rq, 
{\it Annali di Matematica Pura ed Applicata} {\bf 84}, 33-59 (1970).

\bibitem{synge} J.~L.~Synge, {\it Relativity: The Special Theory}, (North-Holland, Amsterdam, 1956), 272.

\bibitem{Aguirregabiria} J.~M.~Aguirregabiria, A.~Hern\'{a}ndez, and M.~Rivas, 
\lq\lq The Li\'{e}nard-Wiechert potential and the retarded shape of a moving sphere\rq\rq,
{\it Am. J. Phys.} {\bf 60} (7), 597-599 (1992).

\bibitem{Griffiths4ed} David~J.~Griffiths, {\it Introduction to Electrodynamics, 4th ed.},  (Pearson, Boston, 2013), 452.


\end{thebibliography}
\end{document}